\documentclass[
    ,final            
  ]
  {aipproc}

\layoutstyle{6x9}

\newcommand{\ve}{\vec}
\newcommand{\de}{\mbox{d}}
\newcommand{\e}{\mbox{e}}
\newcommand{\frc}[2]{\mbox{$\frac{#1}{#2}$}}

\begin{document}

\title{The One-Body and Two-Body Density Matrices of Finite Nuclei 
       and Center-of-Mass Correlations}

                

\author{A.~Shebeko}{
  address={NCS "Kharkov Institute of Physics and Technology", 
           Academicheskaya Str. 1, \\ 61108 Kharkov, Ukraine 
           (shebeko@kipt.kharkov.ua)}
}

\author{P.~Papakonstantinou}{
  address={Institute of Nuclear Physics, T.U.Darmstadt, 
           Schlossgartenstr. 9, D-64289 Darmstadt, Germany 
           (panagiota.papakonstantinou@physik.tu-darmstadt.de)}
}

\author{E.~Mavrommatis}{
  address={University of Athens, Physics Department, 
           Nuclear and Particle Physics Division, 
           Panepistimiopoli, Ilissia, GR-15771 Athens, Greece 
           (emavrom@cc.uoa.gr)}
}

\begin{abstract}
A method is presented 
for the calculation 
of the one-body (1DM) and two-body (2DM) density matrices 
and their Fourier transforms in momentum space, that is 
consistent with the requirement for translational invariance (TI),  
in the case of a nucleus (a finite self-bound system).  
We restore TI by using the so-called fixed center-of-mass (CM) 
approximation for constructing an intrinsic 
nuclear ground state wavefunction (WF) 
by starting from a non-translationally invariant (nTI) WF 
and applying a projection prescription. 
We discuss results 
for the one-body (OBMD) and two-body (TBMD) momentum distributions  
of the $^4$He nucleus calculated with the Slater determinant of the 
harmonic oscillator (HO) orbitals, as the initial nTI WF. 
Effects of such an inclusion of CM correlations are found to be 
quite important in the momentum distributions.

\end{abstract}

\maketitle


\section{Introductory remarks}

The last years the interest in the study of nuclei from both 
experimental and theoretical point of view involves, besides 
the 1DM 
\begin{equation} 
\rho^{[1]}(\vec{r}_1,\vec{r}_{1'}) \equiv A  
\int \Psi^{\ast}(\vec{r}_1,\vec{r}_2,\ldots,\vec{r}_A) 
     \Psi       (\vec{r}_{1'},\vec{r}_2,\ldots,\vec{r}_A) 
{\rm d}^3r_2\ldots {\rm d}^3r_A  
\label{Eq01}  
\end{equation} 
and the OBMD 
\begin{equation} 
\eta (\vec{p}) \equiv \int {\rm e}^{ {\rm i} 
\vec{p}\cdot (\vec{r}_1-\vec{r}_{1'})} 
\rho^{[1]} (\vec{r}_1,\vec{r}_{1'}) 
{\rm d}^3r_1{\rm d}^3r_{1'}  ,  
\label{Eq02}  
\end{equation} 
also the 2DM 
\begin{equation} 
\rho^{[2]}(\vec{r}_1,\vec{r}_2;\vec{r}_{1'},\vec{r}_{2'}) \equiv A(A-1)
\int \Psi^{\ast}(\vec{r}_1,\vec{r}_2,\vec{r}_3,\ldots,\vec{r}_A) 
     \Psi       (\vec{r}_{1'},\vec{r}_{2'},\vec{r}_3,\ldots,\vec{r}_A) 
{\rm d}^3r_3\ldots {\rm d}^3r_A  
\label{Eq03}  
\end{equation} 
and its Fourier transforms, for instance the TBMD 
\begin{equation} 
\eta^{[2]} (\vec{p},\vec{k}) \equiv \int 
{\rm e}^{ {\rm i} 
\vec{p}\cdot (\vec{r}_1-\vec{r}_{1'})} 
{\rm e}^{ {\rm i} 
\vec{k}\cdot (\vec{r}_2-\vec{r}_{2'})} 
\rho^{[2]} (\vec{r}_1,\vec{r}_2;\vec{r}_{1'},\vec{r}_{2'}) 
{\rm d}^3r_1{\rm d}^3r_{1'}    
{\rm d}^3r_2{\rm d}^3r_{2'} .  
\label{Eq04}  
\end{equation} 
The above quantities provide among others information on the 
short-range correlations (SRC) in nuclei. 

A prominent role towards the experimental investigation of the
2DM and related quantities is played by the study of 
electromagnetically induced two-nucleon emission 
$(\gamma , NN)$, $(e,e'NN)$,  
carried out with high precision in photon facilities 
(ELSA, MAMI) and electron accelerators with high energy 
100\% duty-cycle beams (Jlab, MAMI) \cite{BGG03,Sta04}. 

One of the theoretical issues still under discussion is a proper 
consideration of TI and the respective 
separation of spurious CM effects.The latter contaminate 
the calculated observables, when the 
independent-particle shell model and theories which take also 
dynamical correlations into account (e.g., Brueckner-Hartree-Fock, 
Variational Monte Carlo) are used, and inhibit the extraction 
of reliable information on the intrinsic properties of nuclei 
directly from the experimental data 
(see, e.g., \cite{EST1973,DOS1975,VNW1998,Schmid} 
and Refs. therein). 

In this context, we prefer to deal with the intrinsic OBMD and TBMD 
and the associated 1DM and 2DM \cite{SPMxx}. They appear naturally when 
calculating the cross sections of nuclear emission in the 
plane-wave impulse approximation, i.e., neglecting the final-state 
interaction effects and the meson-exchange current contributions to 
the electromagnetic interactions with nuclei. 
Eqs. (\ref{Eq01})-(\ref{Eq04}) 
are modified by properly replacing the $A$ space/momentum vectors 
by the respective 
Jacobi coordinates \cite{SPMxx}. 
In particular, we have calculated the expectation value 
for the TBMD  
\begin{eqnarray} 
{\eta}^{[2]}_{\rm int}(\vec{p},\vec{k}) 
&=& A(A-1)
\langle \Psi_{\rm int} | \delta ( {\hat{\vec{p}}}_{A-1} -
\frc{1}{A}\hat{\vec{P}} - \vec{p} )
\delta ( {\hat{\vec{p}}}_A -
\frc{1}{A}\hat{\vec{P}}  - \vec{k} )
| \Psi_{\rm int} \rangle 
\nonumber \\  
 &\equiv& 
A(A-1) 
\langle \Psi_{\rm int} | 
{{\hat{\eta}}^{[2]}_{\rm int}}(\vec{p},\vec{k}) 
| \Psi_{\rm int} \rangle
\label{e:n21}
,
\end{eqnarray}
where $\hat{\vec{P}}=\sum_{\alpha =1}^A\hat{\vec{p}}_{\alpha}$  
is the total momentum operator for a nucleus of $A$ nucleons 
and $\Psi_{\rm int}$ is the intrinsic WF of the ground state. 

\section{Evaluation of the intrinsic OBMD and TBMD}

The intrinsic quantities of interest are determined as the expectation 
values of appropriate $A-$particle operators, that 
depend on the Jacobi variables, 
in 
intrinsic nuclear states (see Eq.~(\ref{e:n21})). 
An algebraic technique \cite{DOS1975,ShG1974} 
is applied for their evaluation, 
based on the Cartesian representation, in which the coordinate 
and momentum operators are linear combinations of the creation 
$\hat{\vec{a^{\dagger}}}$ and destruction $\hat{\vec{a}}$ operators 
for oscillating quanta in the three different space directions 
(see, e.g., \cite{BoM69}). By following a normal ordering procedure, 
these 
intrinsic operators can be reduced to the form: 
an exponential of the set 
$\{\hat{\vec{a^{\dagger}}}\}$ 
times other exponential of the set $\{\hat{\vec{a}}\}$. 
For example, 
in the case of the intrinsic 
TBMD 
we get the representation 
\begin{eqnarray} 
\label{e:n22}
{{\hat{\eta}}^{[2]}_{\rm int}}(\vec{p},\vec{k})
 &=& (2\pi )^{-6} 
\int \de^3\lambda_1 \de^3\lambda_2
\e^{-{\rm i}\vec{p}\cdot\vec{\lambda}_1}
\e^{-{\rm i}\vec{k}\cdot\vec{\lambda}_2}
\hat{E}_{\rm int} (\vec{\lambda}_1 , \vec{\lambda}_2)
,
\\ 
\hat{E}_{\rm int} (\vec{\lambda}_1 , \vec{\lambda}_2) 
&=&  \e^{ -\frac{p_0^2\lambda^2}{8} }
\e^{ - \frac{A-2}{A} \frac{p_0^2\Lambda^2}{2} }
\hat{O}_1(\vec{z}) \ldots \hat{O}_{A-2}
(\vec{z}) \hat{O}_{A-1}
(\vec{x_2}) \hat{O}_A(\vec{x_1}),
\label{e:n22b}
\end{eqnarray} 
where 
$\hat{O}_{\alpha}(\vec{y})= 
{\rm e}^{-\vec{y}^{\ast}\cdot\hat{\vec{a}}_{\alpha}^{\dagger}} 
{\rm e}^{-\vec{y}       \cdot\hat{\vec{a}}_{\alpha}          } 
$
($\alpha = 1,\ldots ,A$), 
$\vec{\Lambda}=(\vec{\lambda}_1+\vec{\lambda}_2)/2$, 
$\vec{\lambda}=\vec{\lambda}_1-\vec{\lambda}_2$ 
and 
\[ 
\vec{x_1} = \frc{p_0}{\sqrt{2}} (\frc{A-2}{A} \vec{\Lambda} -
\frc{1}{2} {\ve{\lambda}} )  
\,\, , \,\,
\vec{x_2} = \frc{p_0}{\sqrt{2}} (\frc{A-2}{A}
\vec{\Lambda} + \frc{1}{2} {\ve{\lambda}} ) 
\,\, , \,\,  
\vec{z} = - \sqrt{2} \frc{p_0}{A} \vec{\Lambda}
 , 
\] 
with 
$p_0$ the oscillator parameter in the momentum space.
The Tassie-Barker-type factors in Eq.~(\ref{e:n22b}) 
appear in a model-independent way, i.e. they 
result from the intrinsic operator structure and do not depend 
on the intrinsic WF $\Psi_{\rm int}$, which is yet to be determined. 

In the following, we use the intrinsic unit-normalized WF 
constructed from a given nTI WF $\Psi$, 
following Ernst-Shakin-Thaler (EST) 
(fixed-CM approximation) \cite{EST1973},  
\begin{equation}
\label{e:esti}
|\Psi_{\rm int}^{\rm EST} \rangle =
{(\vec{R}=0 |\Psi\rangle }  /
{ [{\langle \Psi | \delta (\hat{\vec{R}}) |\Psi\rangle }]^{1/2} }
\label{Eq08} 
 ,
\end{equation}
where $ | \vec{R}) $ is an eigenvector of  the CM operator
$\hat{\ve{R}} = A^{-1} \sum_{\alpha =1}^A \hat{\ve{r}}_{\alpha}
$ and $\hat{\vec{R}}|\vec{R}=0)=0$.
Here the bracket $|~)$ is used to represent a vector in the space 
of the CM coordinates only. 

\section{Calculation within the Independent Particle Model: 
Application to $^4{\rm He}$ and discussion} 

One can show that if $|\Psi\rangle$ is the Slater determinant 
$|{\rm Det}\rangle$ composed of single-particle (s.p.) orbitals 
$\phi_i(\alpha )$ ($\alpha=1,\ldots ,A$), then the evaluation of the 
distribution $\eta_{\rm EST}^{[2]} (\vec{p},\vec{k})$ 
is reduced to the evaluation of the matrix element 
\[ \langle {\rm Det} |
\hat{O}_1(\vec{z'})
\cdots
\hat{O}_{A-2}(\vec{z'})
\hat{O}_{A-1}(\vec{x'}_2)
\hat{O}_A(\vec{x'}_1)
|{\rm  Det} \rangle 
\label{e:53b}
=\langle {\rm Det}(-\vec{x'}_1,-\vec{x'}_2 ,-\vec{z'} ) |
 {\rm Det}(\vec{x'}_1 ,\vec{x'}_2 ,\vec{z'} ) \rangle
 , \]
that depends on some new complex vectors 
$\vec{x'}_1,\vec{x'}_2,\vec{z'}$~\cite{SPMxx}. 
The OBMD is evaluated in a similar way.  
The Slater determinant $|{\rm Det}(\vec{x'}_1,\vec{x'}_2,\vec{z'})\rangle $ 
is deduced from the original $|{\rm Det}\rangle$ via a substitution 
$|\phi_i(\alpha )\rangle \rightarrow \hat{O}_{\alpha}|\phi_i(\alpha )\rangle$. 
Further analytical evaluations are simplified with the HO orbitals. 
For example, in the simplest case of the $0s^4$ configuration, 
which we encounter in the $^4$He nucleus, one can easily see that 
the matrix element is equal to unity since the s.p. state 
$|0s\rangle$ coincides with the vacuum state of the Cartesian 
representation, viz., $\hat{\vec{a}}|000\rangle=0$, so that 
$\exp{\{\hat{\vec{a}}\}}|000\rangle = |000\rangle$. 
Moreover, the results obtained in this case are independent of the 
projection treatment used (see Eq.~(\ref{Eq08})).  

In this way we have obtained the CM-corrected OBMD of $^4$He  
\begin{equation} 
\eta_{\rm EST}  (\vec{p}) = 
\eta_{\rm EST} (p) = 
4\frac{4^{3/2}b_{\rm cm}^3}{3^{3/2}\pi^{3/2}}
\e^{-\frac{4}{3}p^2b_{\rm cm}^2}
\label{Eq09}
\end{equation} 
vs. the OBMD in the HO model without CM corrections 
\begin{equation} 
\eta_{\rm sp }  (\vec{p}) = 
\eta_{\rm sp } (p) = 
4\frac{b_0^3}{\pi^{3/2}}
\e^{-p^2b_0^2}
\end{equation} 
and the CM-corrected TBMD of $^4$He 
\begin{equation} 
\eta^{[2]}_{\rm EST} (\vec{p},\vec{k}) =
12 \frac{2^{3/2}b_{\rm cm}^6}{\pi^3}
\e^{-\frac{3}{2}p^2b_{\rm cm}^2}
\e^{-\frac{3}{2}k^2b_{\rm cm}^2}
\e^{-\vec{p}\cdot\vec{k}b_{\rm cm}^2}
\end{equation} 
vs. the TBMD in the simple HO model \cite{PMK2003} 
\begin{equation} 
\eta^{[2]}_{\rm sp } (\vec{p},\vec{k}) =
12 \frac{b_0^6}{\pi^3}
\e^{-p^2b_0^2}
\e^{-k^2b_0^2} .  
\label{Eq12}
\end{equation} 
The value for $b_0$ ($b_{\rm cm}=\sqrt{4/3}b_0$) 
is obtained by equating the charge rms radius of $^4$He, 
$r^2_{\rm rms}=\frac{3}{2}b_0^2+b_p^2$ in the simple HO model  
($r^2_{\rm rms} = \frac{3}{2}\frac{A-1}{A}b_{\rm cm}^2+b_p^2$ 
in the HO model with CM corrections), 
with its experimental value 
($r_{\rm rms}=1.67$~fm) and by taking 
the proton rms radius $b_p$ equal to 0.8~fm. 
We find $b_0=1.197$~fm and $b_{\rm cm}=1.382$~fm.


From Eqs.~(\ref{Eq09})-(\ref{Eq12}) 
a shrinking of the distribution 
$\eta_{\rm EST} (p)$ 
($\eta_{\rm EST}^{[2]}(\vec{p},\vec{k})$) 
with respect to 
$\eta_{\rm sp } (p)$ 
($\eta_{\rm sp }^{[2]}(\vec{p},\vec{k})$) 
follows, 
i.e.,  
each of these distributions, 
after being CM-corrected, increases in its central 
but decreases in its peripheral region. 
More exactly, in the case of the TBMD this effect is related to the 
two-dimensional surface given by the function 
$\eta (p,k,\cos{\gamma})\equiv\eta_{\rm EST}^{[2]}(\vec{p},\vec{k})$ 
of the variables $p$ and $k$ at each value of the angle $\gamma$ 
between the vectors $\vec{p}$ and $\vec{k}$. 
As shown in  \cite{DOS1975}, the shrinking of the OBMD 
plays an essential role in getting a fair treatment of the data on the 
inclusive electron scattering in the GeV region. 
Another prominent feature of $\eta_{\rm EST}^{[2]}(\vec{p},\vec{k})$ 
is its asymmetry due to the $\gamma-$dependence. 
Fig.~1 demonstrates these changes for 
$\vec{k}=k_p\hat{p}$, where $k_p$ is positive (negative) for 
$\vec{k}$ in the same (opposite) direction with respect to $\vec{p}$. 
The HO2 curves correspond to the calculations in Ref.~\cite{PMK2003} 
($b_0$ replaced by $b_{\rm cm}$ in Eq.~(12)). 
Among the evident quantitative changes we observe the shift of the 
peak from $k_p=0$ towards negative $k_p$'s, for $p\neq 0$, due to 
a specific correlation induced by the CM fixation. 

In Ref.~\cite{PMK2003} the dimensionless quantity 
\begin{equation}
\xi (\vec{p},\vec{k})
\equiv
\eta^{[2]}(\vec{p},\vec{k})/\eta (\vec{p})\eta (\vec{k})
\end{equation}
was introduced as a measure of different correlations. 
In the complete absence of correlations $\xi$ should be equal to 
$1-1/A$. For a finite, self-bound interacting fermion system, deviations 
of $\xi$ from the above value is a measure of CM and(or) statistical and(or) dynamical 
correlations. 
For the nucleus $^4$He in the simple HO 
(where $A$ equals the level degeneracy of the only occupied state, 
thus statistical correlations are not active), 
\begin{equation}
\xi = 1-1/A = 0.75 
\end{equation} 
for all $\vec{p}$ and $\vec{k}$. 
After fixing the CM, $\xi$ depends on $p$, $k$ and $\gamma$ 
\begin{equation}
\xi_{\rm EST}  (\vec{p},\vec{k})
= 0.89493
\e^{-\frac{1}{6}p^2b_{\rm cm}^2}
\e^{-\frac{1}{6}k^2b_{\rm cm}^2}
\e^{-\vec{p}\cdot\vec{k}b_{\rm cm}^2} . 
\end{equation}
In Fig.~2 $\log_{10}{\xi}$ is plotted as a function of $\cos{\gamma}$ for 
selected values of $p$ and $k$. 
$\xi$ is significantly reduced in forward angles. 
The EST TBMD favors momenta of 
opposite directions as compared to the product of the two OBMD. 
The same holds for the TBMD of $^4$He if Jastrow-type SRC 
are included, as in Ref.~\cite{PMK2003} 
(see the corresponding $\xi$ for the case $p=k=4$~fm, plotted 
in Fig.~2 (dotted line)). 
It is anticipated that within the EST approach additional corrections 
due to SRC will appear 
at high values of $p$ and/or $k$ and that they 
will be larger when $\vec{p}$ and $\vec{k}$ are antiparallel. 
  
\section{Prospects} 
The method presented here is sufficiently flexible to be applied to a 
combined consideration of the CM and SRC. The latter, being 
introduced by means of Jastrow correlations, do not violate the TI. 
We are also planning to extend our elaborations to other $Z=N$ 
light-medium nuclei. 
This general formalism can be 
helpful in studying other two-body and many-body quantities. 


\begin{figure}
 \mbox{} \hspace{-0.45\textwidth} \includegraphics[height=.21\textheight]{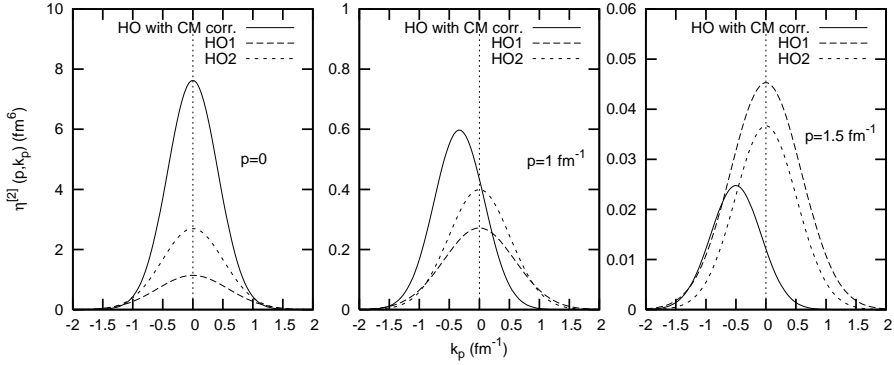}
\caption{%
The TBMD of $^4${He} for $\vec{p}\parallel \vec{k}$
in the HO model with and without 
CM correlations 
(HO1: $b =b_0$; HO2: $b =b_{\rm cm}$).
} 
\end{figure}

\begin{figure}
 \includegraphics[height=.175\textheight]{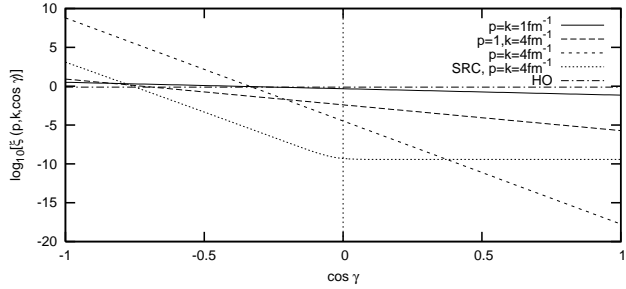}
\caption{%
The quantity $\log_{10}\xi$ of $^4${He} for
the indicated values of $p$ and $k$
as a function of $\cos{\gamma}$, where
$\gamma$ is the angle between $\vec{p}$ and $\vec{k}$.
Full, long and short-dashed lines: with CM correlations;  
dot-dashed: HO model without CM correlations; dotted: including SRC 
but not CM correlations ($b = b_{\rm cm}$).} 
\end{figure}


\bigskip\noindent  
This research has been supported in part by the 
University of Athens under Grant No 70/4/3309 
and by the Deutsche Forschungsgemeinschaft within 
SFB 634. 




\end{document}